\documentclass[twocolumn,amssymb,amsmath,
nofootinbib,tightenlines,showpacs,floatfix,
superscriptaddress,aps,prb]{revtex4-1}
\usepackage{amsfonts,amsbsy,graphicx,bm,color,txfonts,dcolumn,times,bbm}

\begin{document}
\def\bar{\begin{eqnarray}}
\def\ear{\end{eqnarray}}
\def\beq{\begin{equation}}
\def\eeq{\end{equation}}
\newcommand{\degrees}{\ensuremath{^{\circ}}}
\newcommand{\bsigma}{\mbox{\boldmath$\sigma$}}
\newcommand{\bnabla}{\mbox{\boldmath$\nabla$}}
\newcommand{\identity}{\mathbbm{1}}
\newcommand{\ttensor}[1]{\overline{\overline{#1}}}
\newcommand{\nsn}{$N$$S$$N$ }
\newcommand{\fsf}{$F$$S$$F$ }
\newcommand{\sns}{$S$$N$$S$ }
\newcommand{\sfs}{$S$$F$$S$ }
\newcommand{\sffs}{$S$$F$$F$$S$ }
\newcommand{\sfffs}{$S$$F$$F$$F$$S$ }
\newcommand{\sfsfs}{$S$$F$$S$$F$$S$ }
\newcommand{\sfsffs}{$S$$F$$S$$F$$F$$S$ }
\newcommand{\sffsfs}{$S$$F$$F$$S$$F$$S$ }
\newcommand{\sffsffs}{$S$$F$$F$$S$$F$$F$$S$ }
\newcommand{\snsfs}{$S$$N$$S$$F$$S$ }
\newcommand{\snsns}{$S$$N$$S$$N$$S$ }
\newcommand{\sffn}{$S$$F$$F$$N$ }
\newcommand{\sff}{$S$$F$$F$}
\newcommand{\ff}{$F$$F$ }
\newcommand{\fs}{$S$$F$ }
\newcommand{\n}{$N$ }
\newcommand{\f}{$F$ }
\newcommand{\s}{$S$ }
\title{Josephson Currents and Spin Transfer Torques in Ballistic
SFSFS Nanojunctions}
\author{Klaus Halterman }
\email{klaus.halterman@navy.mil}
\affiliation{Michelson Lab, Physics
Division, Naval Air Warfare Center, China Lake, California 93555}
\author{Mohammad Alidoust}
\email{phymalidoust@gmail.com}
\affiliation{Department of Physics,
University of Basel, Klingelbergstrasse 82, CH-4056 Basel, Switzerland}
\date{\today}


\begin{abstract}
Utilizing a full microscopic Bogoliubov-de Gennes (BdG) approach,
we study the equilibrium
charge and spin currents in ballistic
 $SFSFS$ Josephson systems, where $F$ is a 
 uniformly magnetized ferromagnet
 and $S$ is a conventional $s$-wave superconductor.
From the spatially varying
spin currents, we also
calculate the
associated equilibrium spin transfer torques.
Through 
variations in the relative phase differences 
between the three $S$ regions,
and magnetization orientations of the ferromagnets,
our study demonstrates
tunability and controllability of the spin and charge supercurrents.
The 
spin transfer torques are shown to
reveal details of the proximity effects
that play a crucial role in
 these  types of hybrid  systems. 
The proposed \sfsfs nanostructure 
is discussed within the context of  a  
superconducting magnetic torque
transistor.
\end{abstract}
\maketitle
\section{Introduction}
Proximity effects inherent to superconducting systems with
inhomogeneous magnetic order presents
a mechanism by which
dissipationless current flow and
the spin degree-of-freedom can both be
effectively coupled and
controlled.\cite{bergeret1,eschrigh1,efetov1,golubov1,buzdin1,shay} 
The important role that proximity effects
play in the static and transport properties of ferromagnetic 
Josephson junctions with $s$-wave 
superconductors is now well established. 
Indeed, the proximity induced damped
oscillatory superconducting correlations within the ferromagnet 
region serves
as a channel for interlayer coupling and spin switching
\cite{bergeret1,eschrigh1,buzdin1,alidoust_sfsfs,waintal_1,halter_tc,halter_gr_tc}.
Proximity induced triplet pairing correlations within the ferromagnetic junction
also  provides 
another avenue for spin transport.\cite{Eschrig_1,Eschrig_2,halter_trplt,halter_trplt2}
Interest in Josephson junctions with  ferromagnetic 
layers has grown due to their possibility as serving as elements in  next generation superconducting computing and
nonvolatile memories,\cite{eschrigh1,efetov1,golubov1,buzdin1} where 
single flux quantum circuits containing 
multiple Josephson junction arrangements  
can improve switching speeds.\cite{giaz1,giaz3,alidoust_mgmtr} 
To  determine whether Josephson structures can serve as viable cryogenic spintronic 
devices, it is crucial to understand the behavior of the spin currents that can flow
in such systems. 
The spin
current flowing into
the ferromagnetic
regions exerts
a torque on the
magnetization
if the current polarization direction is
noncollinear to the local magnetization
in the ferromagnet. In other words, the
spin angular momentum of the polarized current will
be partially transferred to the magnetization in
the \f region.\cite{STT_1,STT_2}
This 
spin transfer torque (STT)  serves as an important
mechanism  in
spintronics devices.\cite{STT_1,STT_2,linder1} 
The STT effect
can cause magnetization
switching for sufficiently large currents without
the need for an
external field. This 
switching aspect provides a unique opportunity
to create and improve fast-switching magnetic random access
memories. \cite{zutic_rmp,fert_rmp,brataas_nat,Bauer_nat} 

\begin{figure}
\centerline{\includegraphics[width=8.40cm,height=6.0cm]{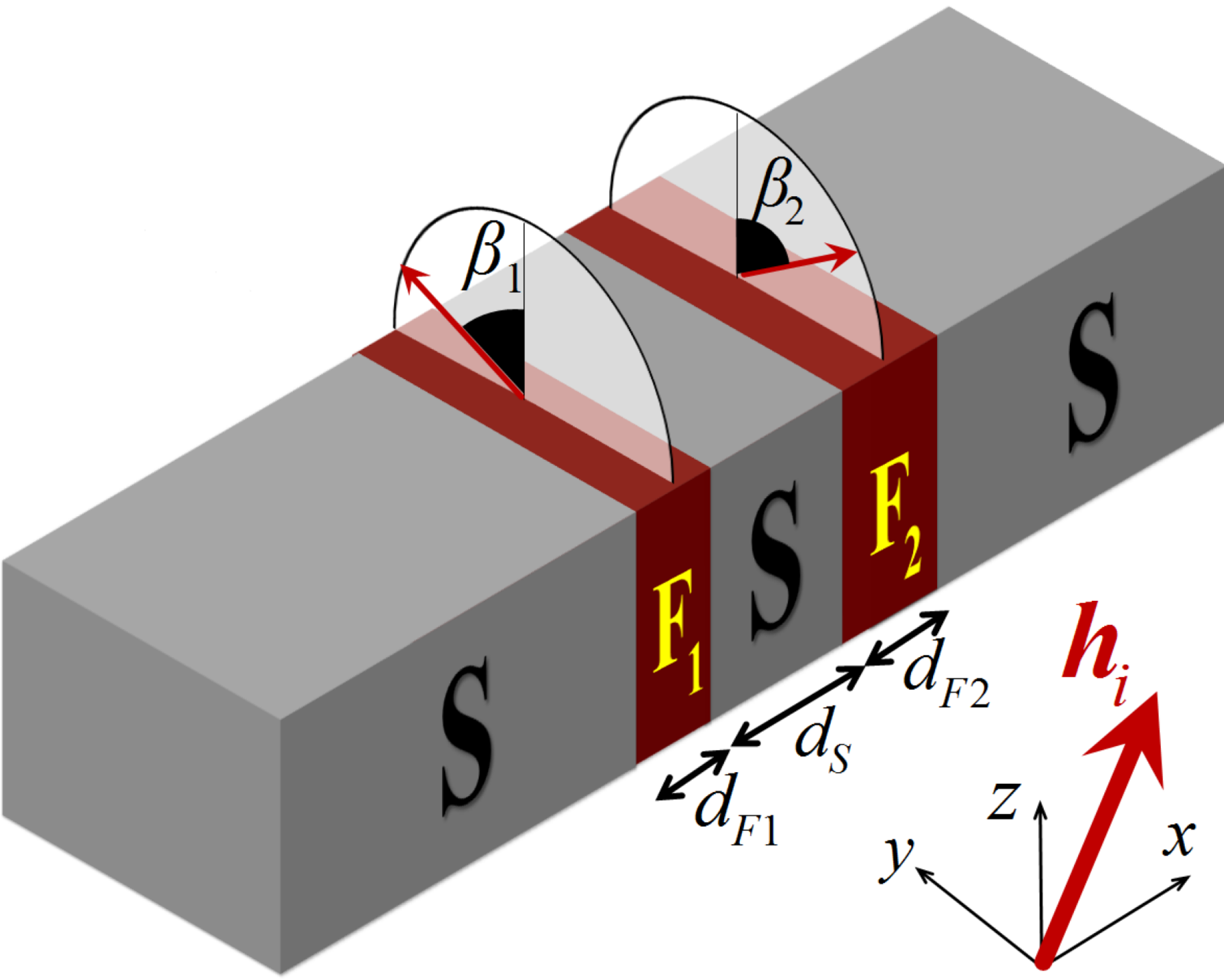}}
\caption{(Color online) Schematic of the \sfsfs ballistic heterojunction
considered in this paper. The 
system is infinite in the $yz$ plane, and
thus
the $x$ axis is normal to the plane of the interfaces.
The middle \s electrode has a thickness  $d_S$ in
the $x$ direction, and is sandwiched between two \f layers with
unequal thicknesses $d_{F1}$ and $d_{F2}$. 
The \f layers are uniformly magnetized and 
their exchange fields are denoted 
by ${\bm h}_i$ (for $i = 1,2$). To simplify 
notation, we have defined the 
magnetization directions via:
${\bm{h}}_i=|{\bm h}_i|(\cos\alpha_i,\sin\alpha_i\sin\beta_i,\sin\alpha_i\cos\beta_i)$.
The \s electrodes
can take arbitrary phases, described by 
$\varphi_L$, $\varphi_M$, and $\varphi_R$
for the
left, middle and right \s electrodes, respectively. 
} 
\label{fig:diagram}
\end{figure}

The recent experimental pursuits of spin-based 
memory technologies involving various arrangements of \sfs
Josephson junctions has rekindled  interest in the realm of ferromagnetic
Josephson
arrays.\cite{alidoust_sfsfs,buzdin_sfsfs,ryaz1,ryaz2,ryaz3,ryaz4,Kupriyanov1}
When a sequence of \sfs
junctions are placed in a series configuration, creating a \sfsfs
type junction shown in Fig.~\ref{fig:diagram},
additional possibilities emerge for the control of
the associated spin and charge
supercurrents.\cite{alidoust_sfsfs,buzdin_sfsfs} 
For example, the triplet components of the supercurrent and total charge transport in diffusive
\sfsfs structures is closely linked to the relative magnetization
orientations, which can directly alter the total charge current flow, causing it to reverse 
direction in the
ferromagnetic layers.\cite{alidoust_sfsfs}
The transport of triplet
supercurrents through the middle \s electrode can be utilized to 
manipulate the magnetic moment of the \f layers in  \sfsfs
hybrids.\cite{buzdin_sfsfs}
The spin-polarized supercurrents in these types of systems 
may also be used to induce
a  STT acting on the magnetization
of a ferromagnet.\cite{waintal_1,waintal_2,Zhao,alidoust_spn} 

At the interfaces between the \f and \s regions in a ballistic
\sfsfs Josephson junction, quasiparticles undergo Andreev and
conventional reflections.\cite{radovic2,radovic1,beenaker2,beenaker1}
Besides the contribution from the
continuum states, the superposition and interference of
the corresponding quasiparticle wavefunctions in the \f regions result in subgap bound states that contribute
to the total current flow between the \s banks. 
By varying the
width of the central \s layer, $d_S$, modifications to the Andreev
bound state spectra can ensue, i.e., by simply decreasing $d_S$,
additional overlap can occur between the subgap bound states in the adjacent \f
regions.\cite{Freyn1,Chang1,Averin}
For \sfsfs
type structures, if the central \s layer is sufficiently thin, i.e.,
$d_S\lesssim \xi$, the proximity effects within the interacting \f regions
result in the Cooper pair amplitude and local density of states
in each \f region being mutually altered. Therefore, magnetization
rotation in a single \f layer can strongly influence the
thermodynamic and quantum transport properties throughout the rest of the 
system.
The
coupling between the different regions can then result in the system
residing in a ground state corresponding to a phase difference of
$\Delta\varphi = \pi$.\cite{alidoust_sfsfs,trifunovic}
The appearance of superharmonic Josephson currents
(with second and higher harmonics: 
$\sin2\varphi$, $\sin3\varphi$, \ldots) were theoretically
predicted to  appear in nonequilibrium and point contact Josephson junctions\cite{2th_hrmnc_1}.
Shortly thereafter, the higher harmonic supercurrents were experimentally observed in  
nonequilibrium situations\cite{2th_hrmnc_2}. 
Also, it was shown theoretically
for a uniform \sfs junction, that the higher 
harmonics can be revealed at the $0$-$\pi$ transition point,
where the first harmonic is highly suppressed
due to the  supercurrent flow reversing
direction at that point.\cite{buzdin1,golubov1} Therefore, the
nonvanishing supercurrent at the $0$-$\pi$ transition point observed
experimentally in Ref.~\onlinecite{ryaz0} was soon attributed to the
presence of higher harmonics\cite{2th_hrmnc_3,buzdin1,zareyan,alidoust_spn}.
Subsequent
works with  ferromagnetic Josephson junctions
demonstrated that the higher harmonics can
naturally arise when varying the location of domain walls\cite{klaus_domain},
and in  ballistic
double magnetic
\sffs junctions, provided that the thickness of the magnetic layers are unequal\cite{trifunovic}. 
Recently, 
evidence of  higher harmonics has been
experimentally observed in Josephson junctions
with spin dependent tunneling barriers.\cite{pure_2th_hrmnc} 

The focus of this paper is to theoretically investigate
proximity effects leading to modified
superconducting correlations and controlled charge
and spin transport
in  \sfsfs ballistic junctions.
We will address a variety of relative magnetization 
orientations, and prescribed superconducting macroscopic phase differences between the
\s terminals. Utilizing a microscopic Bogoliubov-de Gennes
(BdG) approach, we derive the appropriate expressions for
the charge and spin currents and
the corresponding equilibrium spin transfer torques.
The numerical solutions to the BdG equations are employed  to study
the current phase relations, revealing
the emergence of additional harmonics
that depend on the tunable  magnetization profile and other system
parameters.
We demonstrate that our proposed \sfsfs
systems can be considered as a
superconducting magnetic torque 
transistor, where the flow of spin and charge currents can be
tuned by the macroscopic phases of the superconducting leads.
This, in turn, dictates the torques  
acting on the exchange fields of the
\f layers.
Remarkably,
the superconducting phases (in addition to other system
parameters) can effectively switch the torques
acting on the
magnetizations of the \f layers `on' or `off'.
The directions of the torques and charge currents are shown
to not be
related by simple functions of the phase differences or exchange
fields, 
similar to what was observed in  
simpler ferromagnetic 
Josephson junctions involving equilibrium torques.~\cite{waintal_1}
We also find that
when the angle describing the relative in-plane
magnetic exchange field orientations is varied, the torque
tending to align
the two $F$ magnetizations is usually  largest
for relative magnetization angles other than the expected 
orthogonal configurations.
Moreover, we have found that these sequential nanodevices allow for 
detecting pure second harmonics in the current phase relations, depending on the 
system parameters, including the relative magnetization orientations. 
We present a 
study of the crossover between the first and second harmonic in the 
current phase relations and 
consider experimentally feasible situations to observe them. 
This  crossover  is discussed
in the context of the appearance of equal spin triplet correlations with $m = \pm1$ spin projections 
along the spin quantization axis.

The paper is organized as follows. In Sec.~\ref{sec:theory}, we
outline the theoretical approach used and derivation of various
physical quantities investigated, including the supercurrent,
magnetization, spin current, and associated torques. 
We present our results in Sec.~\ref{sec:results}. 
This
section is divided into two-subsections:
Subsect.~\ref{sec:current} 
presents the current phase relations and  the
second harmonic supercurrents that can be generated by calibrating the system parameters. 
Subsect.~\ref{sec:torque}, discusses the associated spin currents and equilibrium spin
transfer 
torques.
Finally, we give concluding remarks in Sec.~\ref{sec:conclusions}.

\section{Theoretical Method} \label{sec:theory}
We begin our methodology by introducing the
Bogoliubov-de Gennes (BdG)
formalism.\cite{Gennes} 
The BdG approach is a convenient
microscopic quantum mechanical technique that allows a complete
investigation into the fundamental characteristics of
the
superconductivity of
ballistic superconducting heterojunctions. 
The microscopic BdG
formalism 
can easily accommodate a broad range of magnetic exchange
field strengths and profiles, including  the half-metallic limit where
the magnitude of the exchange field and the
Fermi
energy, $\varepsilon_F$,  are the same\cite{klaus}. 
A schematic of the multilayer
configuration that we study is depicted in 
Fig.~\ref{fig:diagram}. 
For this quasi one-dimensional  system,  
physical quantities are invariant
with respect to the $yz$ plane, while the $x$-direction captures the
essential physical characteristics of the
system.
The corresponding spin-dependent 
BdG equations are thus expressed as,
\begin{align}
&\begin{bmatrix}
{\cal H}_0-h_z&-h_x+ih_y&0&\Delta(x) \\
-h_x-ih_y&{\cal H}_0 +h_z&\Delta(x)&0 \\
0&\Delta^*(x)&-({\cal H}_0 -h_z)&-h_x-ih_y \\
\Delta^*(x)&0&-h_x+ih_y&-({\cal H}_0+h_z) 
\end{bmatrix} 
\Psi_n(x) \nonumber \\
&\hspace{0.5in}=\epsilon_n
\Psi_n(x),
\label{bogo}
\end{align}
%
where
$\Psi_n(x)\equiv
(u_{n\uparrow}(x),u_{n\downarrow}(x),v_{n\uparrow}(x),v_{n\downarrow}(x))^\mathbb{T}$,
and 
$u_{n\sigma}$ and $v_{n\sigma}$ are the quasiparticle and
quasihole amplitudes.
The pair potential $\Delta(x)$,
which effectively  scatters
electrons into holes (and vice versa)
is nonzero only in the superconducting electrode regions.
We furthermore  assume that $\Delta(x)$ is piecewise constant in the $S$ regions,
with each $S$ region possessing the same magnitude but possibly different phase.
Thus, within the {\it external} $S$ electrodes,
$\Delta(x)$  takes the form
$\Delta_0e^{i \varphi_L}$ in the left, $\Delta_0 e^{i\varphi_M}$ in the middle, and 
$\Delta_0 e^{i\varphi_R}$ in the 
right \s electrode. 
The combinations of phase differences
involving $\varphi_L$, $\varphi_M$, and $\varphi_R$
results in additional possibilities for supercurrent flow
compared to conventional Josephson junctions comprised of
two superconducting banks.
The
single particle Hamiltonian ${\cal H}_0(x)$ is defined as,
\begin{equation} \label{hoho}
{\cal H}_0(x) = -\frac{1}{2m}
\frac{\partial^2}{\partial x^2}+\varepsilon_\perp -\varepsilon_F + U(x),
\end{equation}
where $\varepsilon_\perp=\frac{1}{2m}(k_y^2+k_z^2)$
 is the quasiparticle
 energy for motion in the invariant $yz$ plane (see Fig.~\ref{fig:diagram}),
and the spin-independent scattering
potential is denoted by $U({x})$.
We represent the
magnetism of the \f layers by a Stoner effective exchange energy
${\bm h}({x})$ which will in general have components in all
$(x,y,z)$ directions. 
Additional  technical details on solving the BdG equations 
for this type of
quasi one-dimensional setup
is given in the Appendix.

Various other types of 
``inverse" proximity effects \cite{sillanpaa,halty, berger}
can also occur
in the vicinity of the $F/S$ contacts, 
whereby 
ferromagnetic order propagates from
one \f layer  
to the other (creating a mutual torque)
 via
 the central  \s layer.
Therefore it is of
interest to determine not only the spatial profile of the magnetization ${\bm m}({x})$
within the $F$ regions, but
also within the central $S$ layer, where the induced magnetization can also 
screen\cite{klaus_emerge} the magnetization of the adjacent ferromagnet.
The complete
spatial profiles of the magnetization 
are determined using the expressions: \cite{wu}
\begin{align}
m_x(x)&=- \mu_B\sum_n \Bigl\lbrace \Bigl(u^*_{n\uparrow}(x)u_{n\downarrow}(x)+u^*_{n\downarrow}(x)u_{n\uparrow}(x)  \Bigr)f_n \nonumber \\
&-\Bigl[v_{n\uparrow}(x)v^*_{n\downarrow}(x)+v_{n\downarrow}(x)v^*_{n\uparrow}(x) \Bigr](1-f_n) \Bigr\rbrace. \label{mx} \\
m_y(x)&=- i\mu_B\sum_n \Bigl\lbrace \Bigl(u_{n\uparrow}(x)u^*_{n\downarrow}(x)-u_{n\downarrow}(x)u^*_{n\uparrow}(x)  \Bigr)f_n \nonumber \\
&+\Bigl[v_{n\uparrow}(x)v^*_{n\downarrow}(x)-v_{n\downarrow}(x)v^*_{n\uparrow}(x) \Bigr](1-f_n) \Bigr\rbrace. \label{my} \\
m_z(x)&=-\mu_B\sum_n \Bigl\lbrace \Bigl[|u_{n\uparrow}(x)|^2 - |u_{n\downarrow}(x)|^2 \Bigr]f_n \nonumber \\
&+  \Bigl[|v_{n\uparrow}(x)|^2 - |v_{n\downarrow}(x)|^2 \Bigr](1-f_n) \Bigr\rbrace, \label{mz}
\end{align}
where
$\mu_B$ is the Bohr magneton, and $f_n$ the Fermi function.

The Josephson effect leads
to many possibilities for charge supercurrent transport in  $SFSFS$ junctions.
This experimentally accessible phenomenon  is now well
understood, with the primary driving mechanism being 
the difference between the macroscopic phases of two
\s banks, separated by a weak link.\cite{Likharev,Gennes,josephson}
When there are  three coupled \s banks, the situation becomes more complicated,
and  the 
dissipationless charge current 
depends on  various  combinations
of the phase differences in addition to the other geometric and material properties of the system.
When computing the supercurrent flowing in the $x$ direction,
we 
express
the Josephson
current  in terms of the quasiparticle amplitudes:~\cite{Gennes,klaus_domain}
\begin{align}
\label{cur}
{J_x}({x})&=\frac{2e}{m}\sum_{n}{\rm Im} \Biggl \lbrace f_n\Bigl[u_{n
\uparrow} { \frac{\partial u^{*}_{n \uparrow} }{\partial x}}+ u_{n\downarrow} { \frac{\partial u^{*}_{n \downarrow} }{\partial x}}\Bigr] \\ \nonumber
&+ 
\left(1-f_n\right)\Bigl[v_{n \uparrow}{\frac{\partial v^{*}_{n \uparrow} }{\partial x}} +v_{n \downarrow}{\frac{\partial v^{*}_{n \downarrow} }{\partial x}}\Bigr]
\Biggr\rbrace.
\end{align}

Similarly, following the approach outlined in the Appendix, we can also write the 
spin current  $S_\sigma$ with spin $\sigma$ flowing along the $x$ direction
 in terms of the quasiparticle amplitudes:
\begin{align}
&{S}_x(x) = -\frac{i}{2m}\sum_n \Biggl\lbrace f_n\Bigl[u_{n\uparrow}^* \frac{\partial u_{n \downarrow}}{\partial x}+
u_{n\downarrow}^* \frac{\partial u_{n \uparrow}}{\partial x}-
u_{n\downarrow} \frac{\partial u^*_{n \uparrow}}{\partial x}-
u_{n\uparrow}\frac{\partial u^*_{n \downarrow}}{\partial x} \Bigr ] \nonumber  \label{sx} \\
&-(1-f_n)
\Bigl[v_{n\uparrow} \frac{\partial v^*_{n \downarrow}}{\partial x}+
v_{n\downarrow} \frac{\partial v^*_{n \uparrow}}{\partial x}-
v^*_{n\uparrow} \frac{\partial v_{n \downarrow}}{\partial x}-
v^*_{n\downarrow} \frac{\partial v_{n \uparrow}}{\partial x} \Bigr ] \Biggr\rbrace, \\
&{S}_y(x) = 
-\frac{1}{2m}\sum_n \Biggl\lbrace f_n\Bigl[u_{n\uparrow}^* \frac{\partial u_{n \downarrow}}{\partial x}-
u_{n\downarrow}^* \frac{\partial u_{n \uparrow}}{\partial x}-
u_{n\downarrow} \frac{\partial u^*_{n \uparrow}}{\partial x}+
u_{n\uparrow}\frac{\partial u^*_{n \downarrow}}{\partial x} \Bigr ] \nonumber \\
&-(1-f_n)
\Bigl[v_{n\uparrow} \frac{\partial v^*_{n \downarrow}}{\partial x}-
v_{n\downarrow} \frac{\partial v^*_{n \uparrow}}{\partial x}+
v^*_{n\uparrow} \frac{\partial v_{n \downarrow}}{\partial x}-
v^*_{n\downarrow} \frac{\partial v_{n \uparrow}}{\partial x} \Bigr ]\Biggr\rbrace, \\
&{S}_z (x)=
-\frac{i}{2m}\sum_n \Biggl\lbrace f_n\Bigl[u_{n\uparrow}^* \frac{\partial u_{n \uparrow}}{\partial x}-
u_{n\uparrow} \frac{\partial u^*_{n \uparrow}}{\partial x}-
u^*_{n\downarrow} \frac{\partial u_{n \downarrow}}{\partial x}+
u_{n\downarrow}\frac{\partial u^*_{n \downarrow}}{\partial x} \Bigr ] \nonumber \label{sz} \\
&-(1-f_n)
\Bigl[-v_{n\uparrow} \frac{\partial v^*_{n \uparrow}}{\partial x}+
v^*_{n\uparrow} \frac{\partial v_{n \uparrow}}{\partial x}+
v_{n\downarrow} \frac{\partial v^*_{n \downarrow}}{\partial x}-
v^*_{n\downarrow} \frac{\partial v_{n \downarrow}}{\partial x} \Bigr ]\Biggr\rbrace, 
\end{align}
where the sums for the currents above are in principle taken over  all quasiparticle states.

\section{Results and discussions} \label{sec:results}
We focus here on the low temperature regime, with $T/T_c = 0.001$,
where $T_c$ is the critical temperature of the corresponding bulk 
$S$ material.
For simplicity we set $\varphi_L=0$, $\varphi_M=\varphi/2$, and $\varphi_R=\varphi$,
for the  phases of the
left, middle, and right $S$ terminals, respectively.
Thus, a phase difference of $\varphi/2$ is maintained across each $S$ electrode.
The spatial variables are normalized in terms of 
the  Fermi wavevector, including
the BCS zero-temperature coherence
length, $\xi_0$,  set to $k_F \xi_0 = 100$, and
the dimensionless position $X$,
written as $X=k_{F} x$.
For each physical quantity studied,
a broad range of 
central $S$ widths will be considered.
We  assume that the ferromagnets are similar 
materials with identical exchange field strengths, i.e., 
$|{\bm{h}}_1| = |{\bm h}_2|=h$,
set to the representative value of $h/\varepsilon_F = 0.1$.
To create favorable conditions for equal-spin triplet generation,~\cite{trifunovic}
the ${F_1}$ and ${F_2}$ regions have highly asymmetric widths,
with $d_{F1}=0.1\xi_0$, and $d_{F2}=3.8\xi_0$,
so that $d_{F1} \ll d_{F2}$.

\subsection{Josephson Charge Supercurrent}\label{sec:current}
We begin with a  discussion of the supercurrent charge transport
by 
solving the microscopic BdG equations (Eq.~(\ref{bogo})) 
over a broad range of energies
and then summing the corresponding quasiparticle amplitudes and energies according to the
expression given by Eq.~(\ref{cur}).
The charge current also can be obtained by minimizing the free energy  
with respect to the 
appropriate superconducting phase differences.\cite{Gennes} 
Our microscopic method 
fully accounts for
bound states that may be generated from quasiparticle trajectories with
large momenta corresponding to in-plane energies
comparable  to $\varepsilon_F$ (shown to be important for $SNS$ junctions\cite{sagi}).
The charge supercurrent is normalized by $J_0 = n e v_F$,
where $v_F$ is the Fermi velocity, $e$ the electron charge, and $n$ is the number
density. 
We focus our attention on supercurrents
flowing through the ferromagnets,
recalling that each of the three 
$S$ regions act as effective sources or sinks in the current.
The pair potential $\Delta(x)$ of course vanishes in the intrinsically nonsuperconducting
$F$ regions.

\begin{figure}
\centerline{\includegraphics[width=9cm]{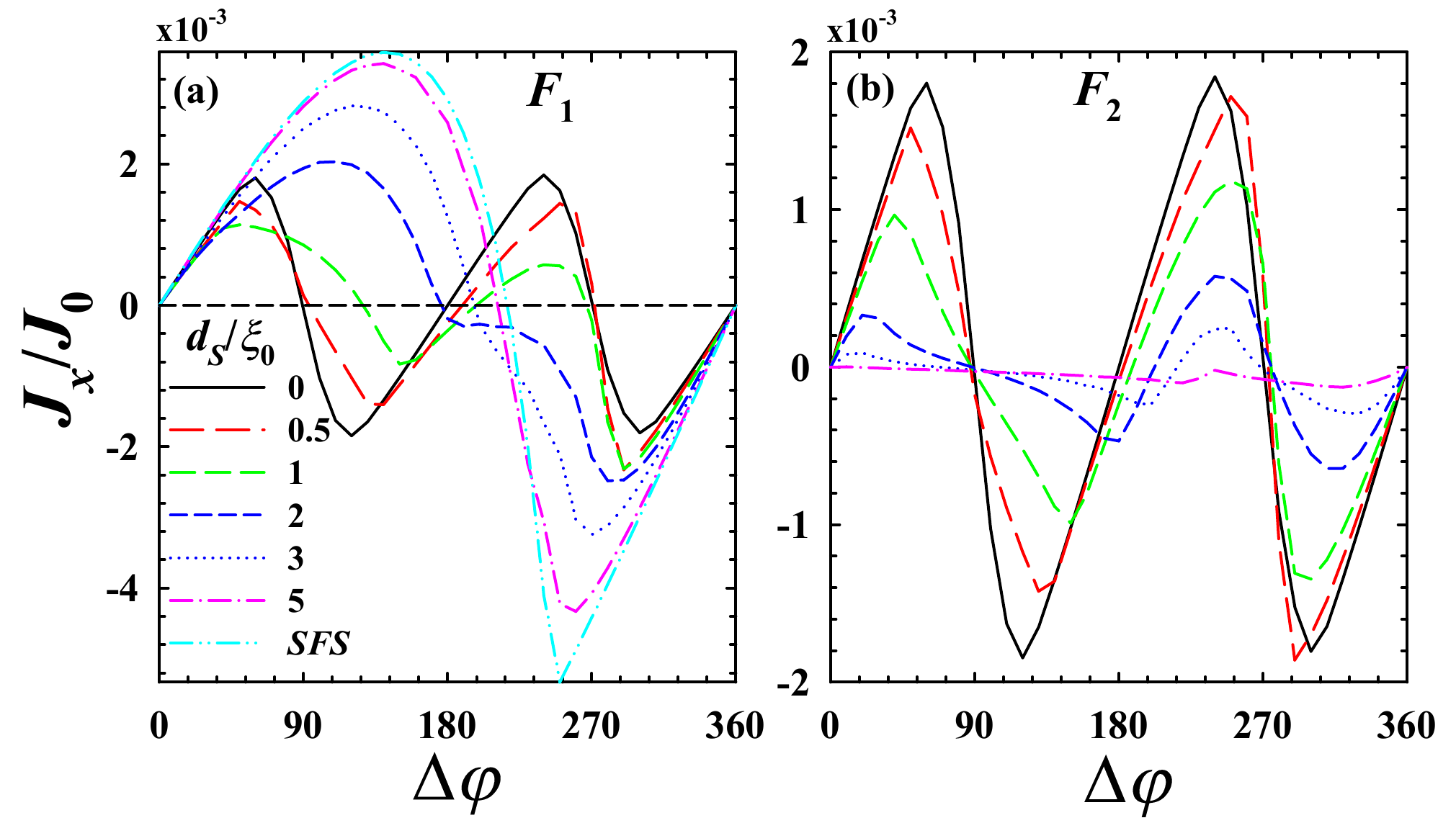}}
\caption{Normalized Josephson current
versus the macroscopic
phase difference of the two outermost electrodes,  $\Delta\varphi$, in a \sfsfs structure with
asymmetric ferromagnet
widths of $d_{F1}/\xi_0=0.1$ and $d_{F2}/\xi_0=3.8$ (see Fig.~\ref{fig:diagram}).
The width of the central $S$ 
electrode, $d_S$, varies as shown in the legend.
The relative exchange fields between the two magnets
is orthogonal with
$\alpha_1=\alpha_2=90^{\rm o}$, $\beta_1=90^{\rm o}$ (along $y$),
and $\beta_2=0^{\rm o}$ (along $z$).
The phase of the middle $S$ electrode takes the value, $\varphi_M=\varphi/2$.
For comparison,  in (a) we show the results for
 a simpler $SFS$ junction having a phase difference $\Delta\varphi/2$ 
and  $F$ width $d_{F1}=0.1\xi_0$.
}
\label{fig1}
\end{figure}
To begin, in Fig.~\ref{fig1}
the supercurrent is shown
as a function of the phase difference, $\Delta \varphi$,
for a wide range of central $S$ electrode widths, $d_S$.
The central $S$ electrode acts as an external current source, and
hence the spatial behavior of the current is piecewise constant in each $F$ region.
Thus, each panel corresponds to the current in a particular ferromagnet (as labeled).
The relative magnetic exchange fields are orthogonal,
with ${\bm h}_1$ directed along $y$  and
${\bm h}_2$
along $z$ (see Fig.~\ref{fig:diagram}).
Two limiting cases are shown: In the first case, the width of the central $S$ layer is zero ($d_S=0$), 
and in the second case,
a large central layer ($d_S=5\xi_0$) is considered.  When there is no
middle $S$ layer,
the current phase relation (CPR) 
is  $\pi$-periodic, with behavior 
consistent with a ballistic $SFFS$ asymmetric double magnetic structure\cite{trifunovic}. 
For $d_S=5\xi_0$, the large $S$ width effectively decouples the two ferromagnets, creating   two isolated $SFS$ junctions
with phase differences $\varphi/2$. 
Thus, in this case
one junction consists of a
thin uniform $F_1$ region sandwiched by two  superconductors,
and its CPR  
 reflects the overall behavior and direction reversal 
that is expected  in
narrow ferromagnetic  $SFS$ Josephson junctions.\cite{golubov1}
The other decoupled $SFS$ junction containing 
$F_2$   (of width $d_{F2}=3.8\xi_0$)
has a  substantially diminished current
due to its much greater width.
Note that there are negligible triplet correlations present
when  $d_S$ is large due to an
effectively uniform magnetization in each
$F$ layer.
For intermediate $S$ layers,
the CPR
evolves from its form in one of these limiting cases,
to
a richer  more complex one
due to the emergence of additional harmonics.
This is due in part to the greater amount of
triplet correlations that are present when the 
$F$ layers possess
orthogonal magnetization configurations.
The
appearance of additional harmonics in the current phase relation 
has been discussed in the diffusive and clean regimes for
simpler ferromagnetic
Josephson junction
structures. \cite{2th_hrmnc_3,alidoust_spn,buzdin1,zareyan}
Note that in Fig.~\ref{fig1}(a),
the  phase corresponding to
the first peak  in the current phase relation,
denoted by the critical phase, $\varphi^*$,
has
$\varphi^* \approx 60^{\rm o}$
for $d_S=0$, and then
gets smaller for $d_S\lesssim \xi_0$ ,
before    
increasing
nearly linearly with $d_S$.
This is contrast to what is observed in Fig.~\ref{fig1}(b), where $\varphi^*$ increases monotonically
with $d_S$.

\begin{figure}
\centerline{\includegraphics[width=9cm]{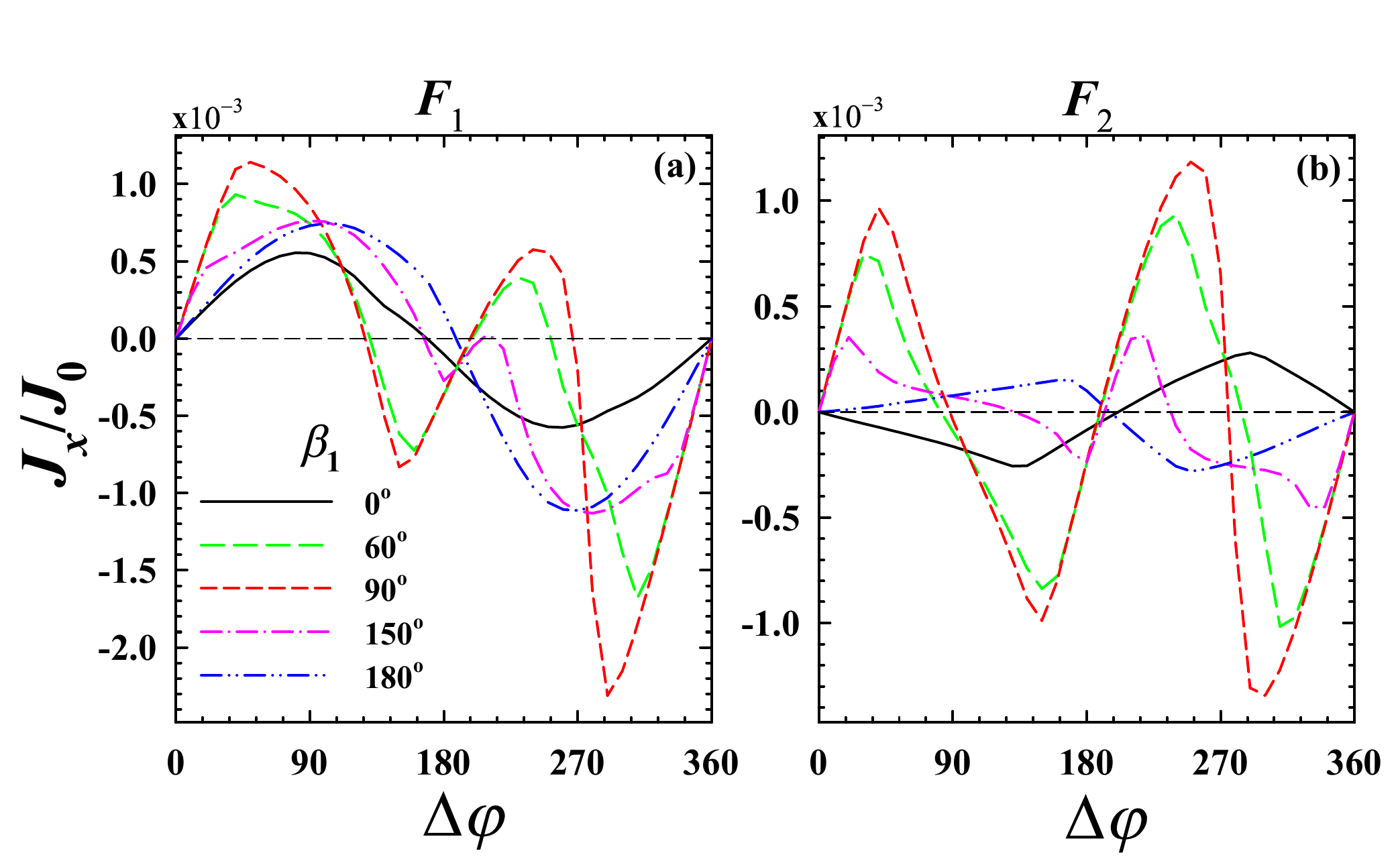}}
\caption{Normalized supercurrent flowing through the ferromagnetic regions vs
$\Delta\varphi$.  The central $S$ layer has width
$d_S=\xi_0$. We consider several magnetization orientations, $\beta_1$ (see legend), inside
the $F_1$ layer of width $d_{F1}=0.1\xi_0$.
The magnetization direction of the larger $F_2$ layer ($d_{F2}=3.8\xi_0$) is strictly 
along $z$, corresponding to $\beta_2=0^\circ$. 
}
\label{fig:JJ_beta2}
\end{figure}
Figure \ref{fig:JJ_beta2} shows the CPRs 
at various magnetization directions $\beta_1$.
The
magnetization  in
 $F_2$ is fixed along the $z$
direction.
The width of the central $S$ layer is now set at $d_S=\xi_0$, and
as stated earlier, its phase has the value $\varphi/2$
to ensure that adjacent superconductors maintain  the same phase
difference.
Examining panel (a), we see that when the relative magnetizations are parallel ($\beta_1=0^\circ$)
or antiparallel ($\beta_1=180^\circ$),
the current exhibits a nearly sinusoidal CPR.
For intermediate
$\beta_1$ leading to noncollinear magnetizations,
higher order harmonics appear in the Josephson current.
Figure  \ref{fig:JJ_beta2}(b) reveals that in the
wider $F_2$ region, collinear orientations result in
regular sawtooth-like patterns in the charge current as $\Delta \varphi$ varies.
The current  in the larger magnet flows in opposite directions
depending on whether the relative  magnetizations are parallel or antiparallel,
in contrast to the narrow $F_1$ segment reported in  (a).
Similarly to what is observed in the narrow $F_1$ region,
we also find more complicated higher order harmonics in describing the current
for misaligned relative magnetizations.

\begin{figure}
\centerline{\includegraphics[width=9cm]{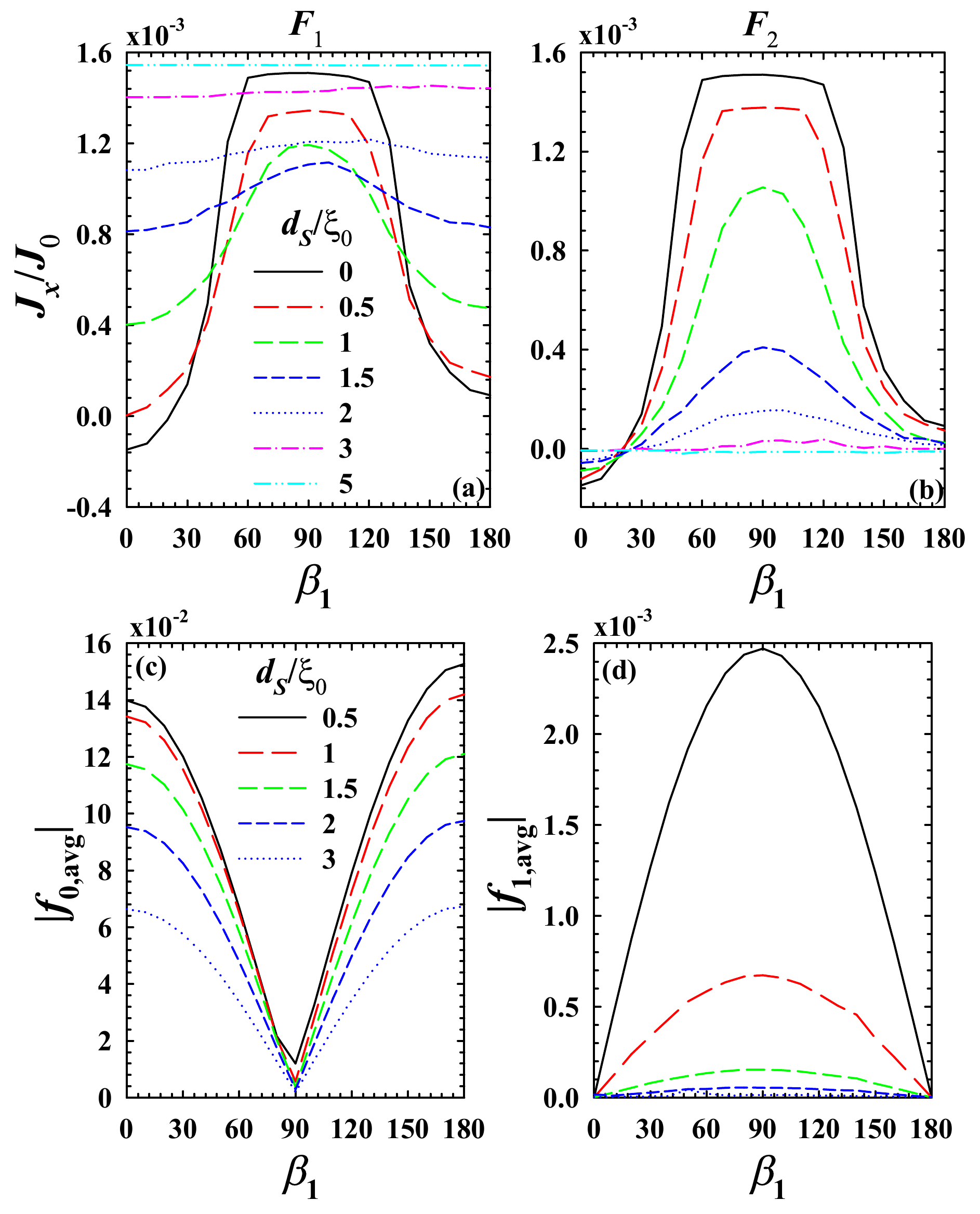}}
\caption
{(a) and (b): Normalized Josephson current
vs the relative in-plane magnetization angle, $\beta_1$
(see Fig.~\ref{fig:diagram}). 
 The exchange fields in the two magnets are parallel when
$\beta_2=0$, and antiparallel when $\beta_2=180^{\rm o}$.
The magnitude of the exchange field 
is set to $h=0.1 \varepsilon_F$.
The    outer $S$ electrodes
have  $\Delta\varphi=45^\circ$.
Five
different $S$ widths are considered, as depicted in the legend.
Panels (c) and (d) show the average
(over the central $S$ region) of the magnitudes of the triplet correlations
vs $\beta_1$.
}
\label{fig3}
\end{figure}
In the broader context of layered  $F/S$ structures,
including ferromagnetic Josephson junctions
and spin valves, misalignment of adjacent
$F$ layer magnetizations will typically generate
equal-spin pairing that is greatest in the 
orthogonal configuration\cite{alidoust_sff,Karminskaya,halt_tc,Fominov_tc}. 
To investigate  the
supercurrent transport properties
when proximity-induced triplet pair correlations are present
in  \sfsfs type structures,
it is instructive  to  investigate
the sensitivity of  $J_x$  to 
relative magnetizations orientation.
Therefore
in Fig.~\ref{fig3}, the Josephson current and triplet correlations are shown in the two ferromagnets as a function
of  exchange field orientation, $\beta_1$.
As $\beta_1$
sweeps between the
parallel ($\beta_1=0^{\circ}$) 
and antiparallel ($\beta_1=180^{\circ}$)
states,
adjacent $S$ electrodes
are, as before, maintained at constant phase difference $\varphi/2$.
Multiple middle $S$ terminal thicknesses are considered (see legend),
with the $d_S=0$ curve  shown for comparison purposes.
As shown in Fig.~\ref{fig3}(a) and (b), 
when magnetic coupling is significant 
(for $d_S\lesssim 2\xi_0$)  the supercurrent is
nonmonotonic and can be highly sensitive to the relative direction of the
magnetic moments 
in the $F$ layers.
When the
magnets have collinear magnetizations, 
the
Josephson current in $F_1$ is often
weaker for the parallel configuration
compared to the antiparallel configuration,
except when there is no central superconductor, or when it is very wide.
The orthogonal state  ($\beta_1=90^\circ$)
however results in the maximal current flow.
Increments in the central electrode thickness can drastically
modify the supercurrent signature.
Eventually for large enough
$d_S$, the magnetic coupling is diminished, 
and variations in $\beta_2$
can no longer affect the current flow. 
The corresponding decoupled $SFS$ junctions
then have  uniform current flow,  that is larger
in the narrow $F_1$ region (panel (a)), and 
in $F_2$ (panel(b)),  becomes negligible due to the larger
width.

The emergence of additional harmonics in the
current phase relations is often correlated with the generation of triplet
correlations that are odd in time~\cite{halter_trplt} or frequency.
As we saw previously in Fig.~\ref{fig:JJ_beta2},
varying the phase difference  $\Delta \varphi$, 
revealed the emergence of
additional harmonics 
as the relative exchange fields went from the
parallel ($\beta_2=0^{\rm o}$) to  orthogonal ($\beta_2 = 90^{\rm o}$)
magnetic
configuration.
To further explore the evolution of 
triplet pairing correlations with magnetic orientations,
in (c) and (d) the spatially averaged
 triplet amplitudes $|f_{0,{\rm avg}}|$ (with spin projection $m=0$), and 
$|f_{1,{\rm avg}}|$ (with spin projection $m=\pm1$) are shown as functions of $\beta_1$.
These quantities are calculated using the expressions:~\cite{halter_trplt}
$f_{0}(x,t) =  {1}/{2}\sum_{n} (f_n^{\uparrow\downarrow}(x)-f_n^{\downarrow\uparrow}(x) ) \zeta_n(t)$,
and 
$f_{1}(x,t)   ={1}/{2} \sum_{n}
(f_n^{\uparrow\uparrow}(x)+f_n^{\downarrow\downarrow}(x))\zeta_n(t)$,
where we define 
$\zeta_n(t) \equiv \cos(\epsilon_n t)-i\sin(\epsilon_n t) \tanh ({\epsilon_n}/{(2T)})$,
and
$f_{n}^{\sigma\sigma'}(x) \equiv u_{n \sigma} (x)v^{\ast}_{n\sigma'}(x)$.
The summations  are in principle over all states.
A representative value of  $\tilde{t}=6$ is used for
the scaled 
relative time, where
$\tilde{t}\equiv \omega_D t$.
The quantization axis in the regions of interest  is 
aligned along the $z$-direction, 
however it is
straightforward to align it
along a different axis
that may coincide with
the local magnetization direction.~\cite{halter_gr_tc}
Comparing Figs.~\ref{fig3}(c) and (d), 
it is evident that the
behavior of the triplet amplitudes as a function of $\beta_1$ is
anticorrelated, with the average $|f_0|$ smallest when
$|f_1|$ peaks at $\beta_1 =  90^\circ$.
Increasing the $S$ width is shown to reduce the $f_0$
amplitudes gradually, however the equal-spin component
$f_1$ drops much more abruptly to negligible values once $d_S$
exceeds $\xi_0$.
Although not shown, the
singlet correlations within $S$ (for all $S$ widths)
were found to not exhibit significant sensitivity 
to changes in $\beta_1$.
This is clearly in sharp contrast to what is observed 
for both triplet components.
Having discussed now  some salient features of 
the charge currents, we now turn our attention
to spin transport and the corresponding equilibrium spin transfer 
torques within the junction region.

\subsection{Spin currents and spin transfer
torques}\label{sec:torque}
Spin-polarized transport quantities are of paramount
importance
when studying \sfsfs type junctions as 
potential components in spintronics 
devices. 
The spin current ${\bm S}$
is a local  quantity 
responsible for the change 
in  magnetizations due
to the flowing of spin-polarized currents.
The main contributor to the equilibrium spin current
and corresponding spin-transfer torque $\bm{\tau}$
is
the
spin-resolved Andreev
bound states \cite{Zhao},
which play the main role in 
torque sensitivity to variations in
$\Delta\varphi$ and $\beta_1$.
Thus, the STT can be a useful probe of the spin degree of freedom
in $S/F$ proximity elements. 
The current that is generated from the macroscopic phase differences in the $S$
electrodes
can become spin-polarized\cite{alidoust_spn,waintal_1,waintal_2,zareyan_3}
when entering one of the ferromagnet
regions. 
A portion of this spin current can then interact with
the other ferromagnet and be absorbed by the local magnetization due to the spin-exchange
interactions.~\cite{wuwu}
Since we are  
considering 
ferromagnets with in-plane magnetic exchange fields,
the only
spin current  that can
flow is  the  out-of-plane component $S_x$. 
This is consistent with the fact that
only $\tau_x$ can exist in equilibrium when
spin currents do not enter or leave the superconducting electrodes.\cite{waintal_1}

As shown in the Appendix, the method used here
to  determine $\bm \tau$
involves 
simply 
calculating the magnetic moment throughout the entire system
and then using,
\begin{align} \label{tau1}
{\bm \tau} = -\frac{2}{\mu_B} {\bm m} \times {\bm h},
\end{align}
where the magnetization components are given in Eqs.~(\ref{mx})-(\ref{mz}).
Equivalently, in the steady state,
one can use
the continuity equation
for the spin current [Eq.~(\ref{scon})]  to determine the torque transferred by
simply evaluating the
derivative of the spin current as a function of position:
\begin{align} \label{sx1}
{\bm \tau}=
\frac{\partial {\bm S}}{\partial x}.
\end{align}
The net flux of spin current $\Delta S_x$ through a certain region
bound by points $x_1$ and $x_2$
is therefore:
\begin{align} \label{sx2}
 \Delta S_x = {\bm S}_{x}(x_2)- {\bm S}_{x}(x_1)=
\int^{x_2}_{x_1} dx  { \tau}_x= \tau_{x,{\rm tot}}.
\end{align}
In other words, the change in spin current
at the interface boundaries  ($x=x_1$ and $x=x_2$)
is equivalent to
the net torque acting within those boundaries.
Either approach, using Eq.~(\ref{tau1}), or Eq.~(\ref{sx1}),
is sufficient to calculate ${\bm \tau}$, as they both
yield precisely the same result.
In the results that follow, we 
calculate the torques using  Eq.~(\ref{tau1}), 
thus avoiding the numerical derivatives that arise when 
using Eq.~(\ref{sx1}).

\begin{figure}
\centerline{\includegraphics[width=8.9cm]{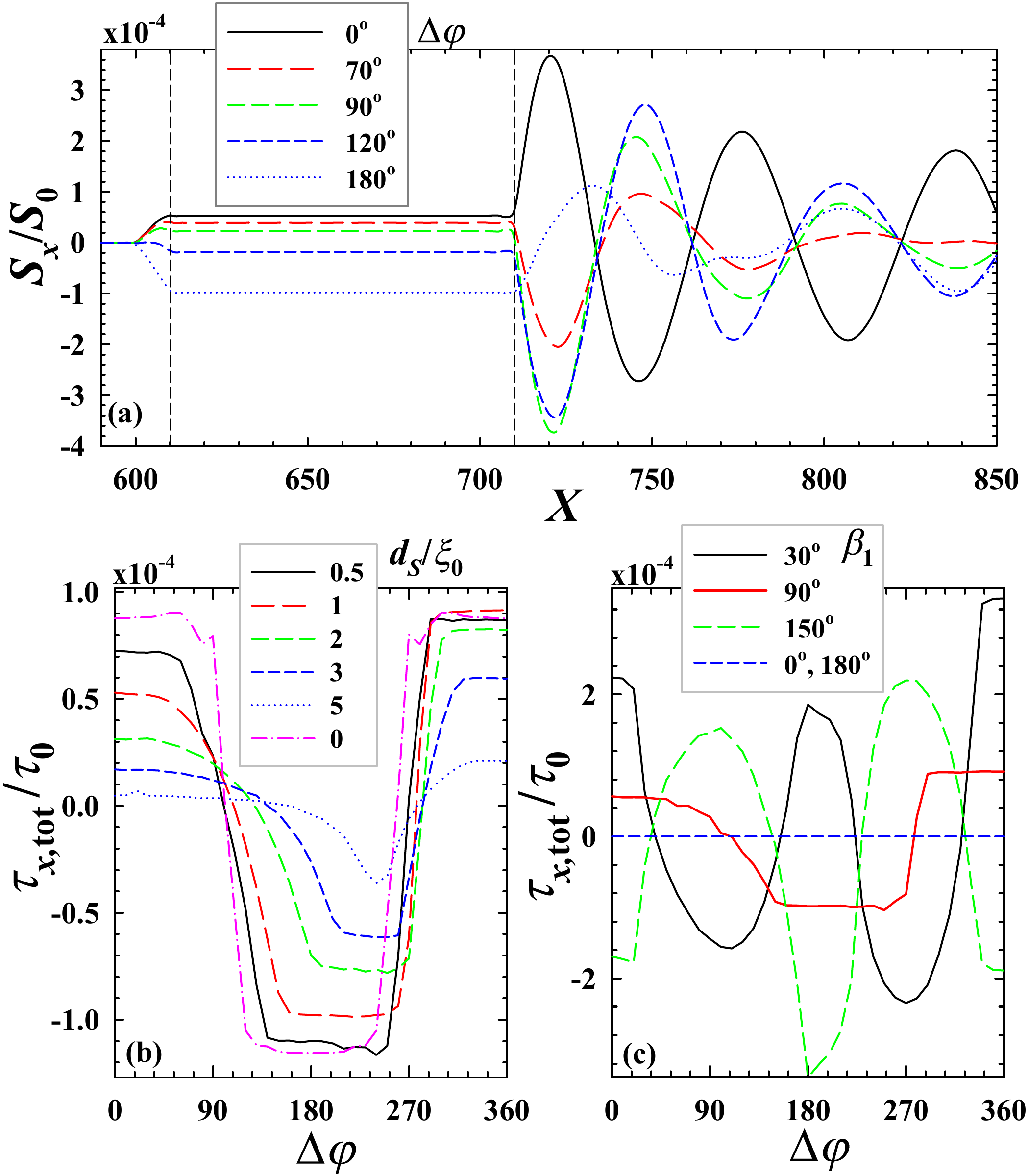}}
\caption
{
(a) Normalized  spin current, $S_x$,
as a function of normalized position $X$. Several phase differences $\Delta\varphi$
are considered.
The interfaces separating each region are
denoted by  vertical dashed lines.
The spin current is conserved in the central $S$ region,
where there is no magnetic exchange interaction.
(b) Normalized total equilibrium torque, $\tau_{x,{\rm tot}}$,
 as a function of the superconducting phase difference $\Delta \varphi$ 
 between the outermost $S$ electrodes.
A wide range of  $d_S/\xi_0$ ratios are considered (see legend).
}
\label{tx_phir}
\end{figure}
We first present 
in Fig.~\ref{tx_phir}(a) 
the $x$-component of the local
spin current, $S_x$, 
normalized
by $S_0\equiv  -\mu_B {\cal N}_F \varepsilon_F/k_F$,
where  ${\cal N}_F$ is the density of
states at the Fermi energy. 
We numerically
calculate $S_x$
by summing the quasiparticle amplitudes and energies using Eq.~(\ref{sx}).
Several different phase differences are studied as shown in the legend,
and
the exchange interactions are
orthogonal: ${\bm h}_1=(0,h_{y1},0)$, and
${\bm h}_2=(0,0,h_{z2})$. 
The central $S$ layers is one $\xi_0$ wide.
The spin current
reveals precise 
spatial behavior of 
the junction 
interlayer magnetic coupling, 
and 
from Eq.~(\ref{sx1}), 
one can deduce the  
corresponding 
local behavior of $\tau_x$. 
In $F_2$, the oscillating spin currents
each have a phase and magnitude 
that can change, depending on $\Delta\varphi$. 
Once the spin current enters the 
$S$ region (bound by the 
dashed vertical lines),
it immediately becomes
conserved,
whereby
there is no
transfer of spin
angular momentum. Since $\partial S_x/\partial x =0$, we
have $\tau_x=0$ in that region.   
Within $F_1$,
the narrow width
limits the extent at which $S_x$ can vary, and consequently 
it undergoes a nearly monotonic decline, before
vanishing within the superconductor. 
Since $S_x=0$ in the outer $S$ electrodes,
the difference $\Delta S_x$ over either ferromagnet is determined by
the value of $S_x$ within the central $S$.
Thus, despite  drastically different local behavior of
$S_x$ in each $F$ region, the net flux of spin current
through either $F_1$ or $F_2$ differs only in sign.
One can then see that by examining 
the value of the conserved $S_x$ in the central $S$ region,
the flux $\Delta S_x$ through e.g., $F_1$ is largest when 
$\Delta\varphi=180^\circ$, and smallest  when 
$\Delta\varphi=120^\circ$.
This observation is consistent 
with (b), where the  total torque (normalized by $\tau_0 \equiv -\mu_B {\cal N}_F\varepsilon_F$)
is shown as a function of $\Delta\varphi$.
We calculate 
$\tau_{x,{\rm tot}}$ 
over the $F_1$ region using Eq.~(\ref{sx2}), although the result
for $F_2$ is trivially obtained, 
since within each of the three $s$-wave $S$ electrodes,
there can be no flux of spin current.
This requires
$\tau_{x,{\rm tot}}$
in $F_1$ to be the exact opposite in $F_2$.
As seen in (b), the net torque can be quite sensitive to the phase difference
$\Delta\varphi$, which when tuned appropriately, can flip direction or 
vanish altogether.
For comparison, 
the $d_S=0$ case is included, which has
symmetric behavior
about  $\Delta\varphi=180^\circ$. 
For most $S$ layer widths considered,  $\tau_{x,{\rm tot}}$  vanishes at
$\Delta \varphi \approx 90^{\rm o}$, and $\Delta \varphi \approx 270^{\rm o}$
before  reversing direction.
Since the middle $S$ terminal has $\varphi_M=\varphi/2$,
the central $S$ electrode 
tends to  asymmetrically distort the supercurrent
about $\Delta \varphi_R=180^\circ$.
Comparing Fig.~\ref{fig1}(a) with Fig.~\ref{tx_phir}(b),
it is evident that the charge supercurrent
$J_x$
is not  simply correlated with 
the flux of spin current  
(or equivalently $\tau_{x,{\rm tot}}$), 
consistent with
previous work\cite{waintal_1}.
The coupling between ferromagnets is 
clearly stronger the thinner the 
$S$ electrodes, where
 $\tau_{x,{\rm tot}}$  is larger and tends to
change less
over a broader range of $\Delta \varphi$,
reflecting a tendency for the magnetization  
to remain fixed 
in place despite
supercurrent variations. 
Eventually however, for sufficient increments in $\Delta\varphi$,
the net torque will abruptly 
reverse direction.

The proposed 
\sfsfs
system 
can 
be 
considered 
as a
type of 
superconducting 
magnetic torque 
transistor, 
where the flow of spin 
and charge currents are
tuned by $\Delta\varphi$.
This, in turn, dictates the torques  
acting on the exchange fields present in the
\f layers.
By minimizing the free energy,~\cite{waintal_1}   
it was shown that
changes in the supercurrent with respect to relative magnetization orientation 
results in a torque  that changes 
with  $\Delta\varphi$, and vice versa.
To underscore the sensitivity of the net torque
to the phase and relative magnetic orientations,
Fig.~\ref{tx_phir}(c) illustrates 
$\tau_{x,{\rm tot}}$ as a function $\Delta\varphi$
for a few  orientation angles, $\beta_1$.
When $\bm m$ and $\bm h$ are collinear, i.e., the two 
exchange field alignments in the ferromagnets are parallel  ($\beta_1=0^\circ$)
or antiparallel ($\beta_2=180^{\circ}$) to one another, 
${\bm m} {\bm \times} {\bm h} = 0$, and hence the net torque is zero (see  Eq.~(\ref{tau1})).
The previous $\beta_2=90^{\circ}$ case in (b) is also shown here.
For noncollinear 
magnetizations, a ``static" torque even exists in the absence of 
a supercurrent ($\Delta\varphi =0$). In this case, the effectively inhomogeneous magnetization
generates a spin current imbalance and torque that tends to align the magnetizations.
When the magnetizations are misaligned,  the
supercurrent can change both
the direction and amplitude of
the  torque\cite{buzdin_sfsfs}. 
In many cases, this effect can be
attributed to the torque that the equal-spin triplet component of
the supercurrent (possessing net spin along the spin quantization axis) 
exerts on the magnetization and tends to rotate it.
As we clearly see from the results presented in Fig.~\ref{tx_phir}(c), 
for thin central superconductors with $d_S\lesssim \xi_0$,
this 
effect
can be quite sensitive to the multiple superconducting phase differences.

\begin{figure}
\centerline{\includegraphics[width=8.8cm]{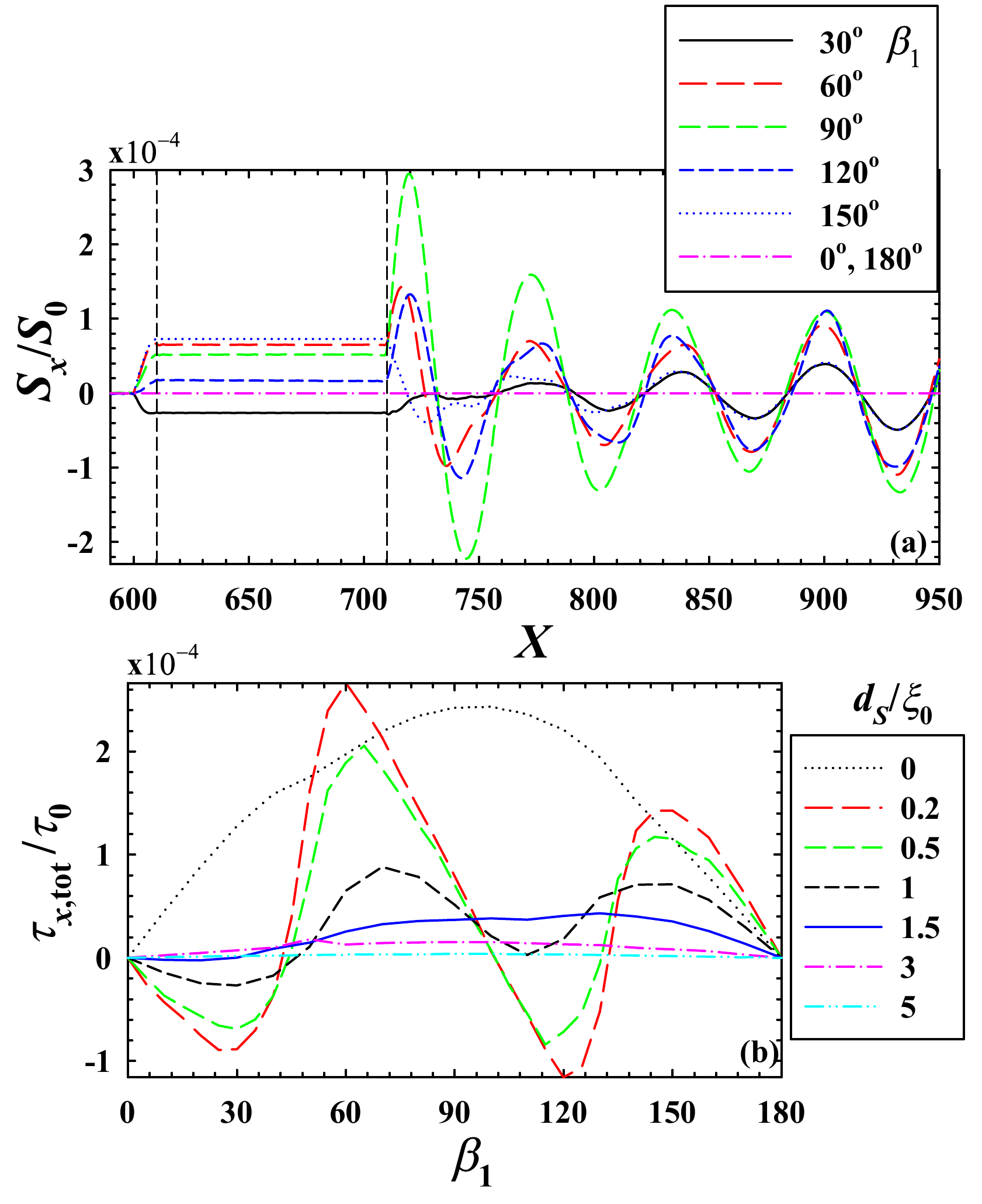}}
\caption
{
(a) The $x$ component of the
spin current $S_x$ flowing throughout a segment of the
\sfsfs  junction, as a function of position $X$.
The magnetic exchange orientation in $F_1$
is varied according to ${\bm h}_1 = h(0,\sin\beta_1,\cos\beta_1)$,
while in
$F_2$ we have:, ${\bm h}_2=(0,0,h)$.
The intermediate $S$ width corresponds to  $d_S/\xi_0=1$.
The  legend
identifies the different angles $\beta_1$ used.
A current is established 
via a phase difference of
$\Delta\varphi=45^\circ$ between the outer $S$ layers.
Spatial variations in $S_x$,
are responsible for any torques present in the
system. 
In (b) the total torque, $\tau_{x,\rm tot}$,
is plotted as a function of $\beta_1$.
The $d_S =0$ reference case is
multiplied by a constant factor for comparison
purposes. Each curve corresponds to
a different $d_S$ as identified in the legend.
}
\label{tx_phi2}
\end{figure}
To investigate further the behavior  of 
 the local spin transport, and total torque 
when varying the ferromagnet orientation angle $\beta_1$,
the
spatial behavior of  spin current throughout the system
is shown in Fig.~\ref{tx_phi2}(a).
The angle $\beta_1$ describes the rotation of the
in-plane magnetic exchange
in $F_1$:
${\bm h}_1= h(0,\sin\beta_1,\cos\beta_1)$.
The magnetic exchange field direction
in $F_2$ does not vary and is directed along $z$: ${\bm h}_2=(0,0,h)$.
Control of the  free-layer magnetization by an external magnetic field
has experimentally been demonstrated
in $S/F$ spin valves \cite{ilya}. 
The rotation angle can also be manipulated 
by STT switching.\cite{Bauer_nat,brataas_nat}
We see that in the $S$ region,
$S_x$ is constant for all angles $\beta_1$,
consistent with
the spin-torque
conservation law [Eq.~(\ref{sx1})]
which states that
any spatial variations of the spin current
must generate
a torque.
The torque thus vanishes in the $S$ region, as it should.
The $F_2$ region again exhibits a spatially modulating spin 
current whose behavior is highly sensitive to 
the particular orientation angle $\beta_1$.
The most rapid changes in the oscillating $S_x$ tends to occur within this
ferromagnet
near the interface with a superconductor.
We also see that the spin current at the interfaces between the ferromagnets and the central $S$ (dashed vertical lines)
varies non-monotonically, changing sign at $\beta_1=30^\circ$, or vanishing altogether when the
magnetizations are collinear ($\beta_1=0^\circ$, or $\beta_1=180^\circ$).
These observations are consistent with  Fig.~\ref{tx_phi2}(b), where 
the total torque $\tau_{x,{\rm tot}}$ is shown as a function of orientation angle $\beta_1$
for several $S$ widths.
We see that $\tau_{x,{\rm tot}}$
vanishes entirely when the two ferromagnets have collinear magnetizations,
corresponding to  parallel ($\beta_1=0^\circ$) or antiparallel ($\beta_1=180^{\circ}$) configurations.
The total torque also has the expected behavior when there is
no middle $S$ terminal ($d_S=0$), peaking when $\beta_1\approx 90^\circ$,
corresponding to  the situation where the torque has the greatest tendency
to align the magnetic moments.
Including a central $S$ layer is seen to introduce
a nontrivial oscillatory behavior in $\tau_{x,{\rm tot}}$
that can cause it to vanish (or change direction)
multiple times when spanning the full $\beta_1$ range.
In effect,  the
angle  $\beta_1$ that was previously responsible for the largest total torque (when $d_S=0$) 
is now the angle at which there is negligible total torque
within the $F$ layers.
Increasing the $S$ thickness of course
reduces the ferromagnetic coupling and hence reduces the magnitude
of the mutual torques.
The misalignment angle  where the maximum torque is
exerted, $\beta^*_1$,
clearly
shifts from near the orthogonal
configuration ($\beta^*_1\approx 90^{\rm o}$) when $d_S=0$
towards
intermediate magnetic configurations corresponding to   $60^\circ\lesssim \beta_1 \lesssim 70^\circ$,
for $d_S\lesssim \xi_0$.

\begin{figure}[t!]
\centerline{\includegraphics[width=8.5cm]{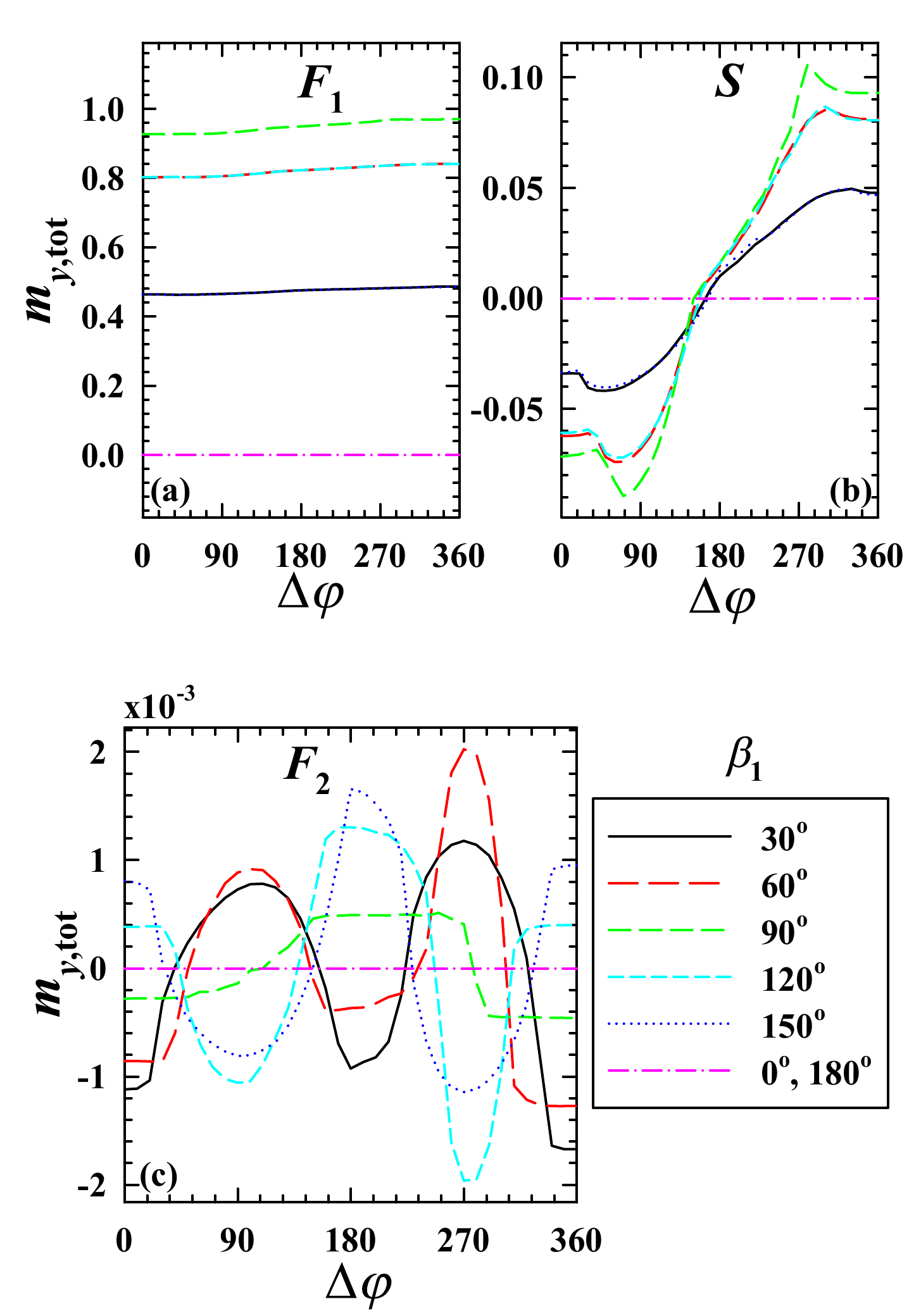}}
\caption
{
The normalized $y$
component of the
 total magnetization $m_{y,{\rm tot}}$, as a function of phase, $\Delta\varphi$,
within the \sfsfs system.
is shown for a few in-plane exchange field
orientations, $\beta_1$ (see legend).
The $F_1$ region has a thickness corresponding to $d_{F1}/\xi_0=0.1$,
$F_2$ has $d_{F2}/\xi_0=3.8$,
and the central layer is one coherence length
wide, i.e., $d_S/\xi_0=1$.
The exchange field has magnitude $h=0.1$.
Each panel (a)-(c) depicts 
a different region where the net magnetization is calculated.
The leakage of magnetism into the central $S$ region
[panel (b)] is clearly visible.
}
\label{mm}
\end{figure}
To examine  the previous behavior of the  total STT
from a different perspective,  
it is beneficial to recall the simple expression, Eq.~(\ref{tau1}),
which shows that
for a given exchange field,
the torque arises entirely from the magnetization, ${\bm m}(x)$.
Thus, it is insightful to study the details of
${\bm m}(x)$, which gives
a measure of the spin polarization in the system responsible
for generating the local spin currents.
The out-of-plane torque is due to
 both in-plane components of the magnetization:
$\tau_x (x)= -(2/\mu_B) [m_y (x) h_z (x) -h_y (x)  m_z (x)]$,
which clearly vanishes outside of the ferromagnet regions
where ${\bm h}_i=0$.
The exchange field in the $F_1$ region
varies in the $yz$ plane, while
in $F_2$,  the only nonzero component to the
exchange field is $h_z$, so that we have simply,
$\tau_x (x)= -(2/\mu_B )h_z (x)  m_y (x)$.
Thus
for a mutual torque to exist in the ferromagnets,  
a $y$-polarized magnetization in
$F_1$ must 
propagate through the central $S$ electrode
and into
the $F_2$ layer,
generating a spin-imbalance.
In Fig.~\ref{mm}(a-c), we illustrate the total magnetization, $m_{y,{\rm tot}}$,
in each region as a function of $\Delta\varphi$.
Here we define the quantity $m_{y,{\rm tot}}$ as the $y$-component of the magnetization
spatially integrated
over each of the three junction regions of interest, and 
normalized by $m_0 \equiv -\mu_B {\cal N}_F$.
Using this normalization, the bulk value of the
magnetization is  equivalent to
the exchange field value of $h/\varepsilon_F=0.1$.
The thicknesses of the 
$F_1$, $F_2$, and $S$ layers are given by $d_{F1}/\xi_0=0.1$,
$d_{F2}/\xi_0=3.8$, and  $d_{S}/\xi_0=1$, respectively.
It is evident that for a wide range of $\Delta\varphi$,
a net
magnetization exists in each junction region,
other than when $\beta_1$ corresponds to relative collinear magnetizations (${\bm h}_1$
directed  along $z$).
In panel (a),  $m_{y,{\rm tot}}$ is approximately constant for each $\beta_1$,
and the overlapping curves at $\beta_1=30^\circ,150^\circ$ reflect the symmetry about
 $\beta_1=90^\circ$, where  the  net magnetization is greatest. 
Also, the net magnetization in $F_1$ is positive for the range of $\beta_1$ shown, since
the magnetic exchange interaction in the $y$ direction
is
$h_{1,y} = h \sin \beta_1$.
Due to proximity effects, 
there is an intrinsic net magnetization in the superconductor that
is present even in the absence of current flow.
As shown in panel (b),  the total  induced magnetization
in the superconductor  is finite at $\Delta\varphi=0^\circ$
and vanishes  at $\Delta\varphi\approx180^\circ$,
where it switches direction.
Finally,  in Fig.~\ref{mm}(c)  the
contribution from the magnetization to 
the total torque observed in Fig.~\ref{tx_phir}
is evident, where the
total magnetization in the wider
$F_2$ region exhibits the same dependence on $\Delta\varphi$
as $\tau_{x,{\rm tot}}$,
differing only in sign.

\section{Conclusions} \label{sec:conclusions}
In conclusion, we have presented a detailed microscopic 
study of the charge and spin supercurrents  that can
exist in  \sfsfs types of Josephson junction hybrids.
The local magnetization profiles were then calculated and employed
for
determining the  equilibrium spin transfer torques.
We also studied the associated  
spin-triplet correlations that
arise in these hybrids.
This was accomplished by 
solving the 
BdG equations
over a
broad range of geometrical and material parameters.
Our investigations revealed how to manipulate and generate 
supercurrents with higher order harmonics by varying the macroscopic 
phases in the superconducting electrodes,
or the relative exchange field  orientations.
Utilizing the spin conservation law, we calculated
the spin transfer torque in these systems, revealing
a number of experimentally viable ways in which
the magnetization can be controlled
in a prescribed fashion.
Our results demonstrate that, depending on the parameters considered,
these types of ballistic systems can support 
supercurrents that can be tuned to contain primarily the  first or second 
harmonics in the current-phase relations.
We studied the $\pi$-$2\pi$ harmonic crossovers and determined the
experimentally desirable conditions in which to reveal the second harmonic
supercurrents in these systems.
We also showed that the equilibrium spin transfer torques can be well
controlled by simply modulating the macroscopic phases of the three
$S$ electrodes in addition to the other system parameters such as the 
sizes of the ferromagnets and
central superconductor electrodes, or the relative magnetization alignments.
These findings are suggestive of a phase-tunable superconducting transistor based on STT switching.

\acknowledgments K.H. is supported in part by ONR and by a grant of
supercomputer resources provided by the DOD HPCMP.
M.A. would like to thank A. Zyuzin for helpful discussions.

\appendix
\section{Numerical procedure for solving the BdG equations}
The numerical procedure used
in calculating the spin and charge currents involves 
first expanding\cite{klaus} the quasiparticle amplitudes 
in terms of a complete set of $N$ basis functions:
\begin{align}
\label{basis}
\psi_{n }(x) = \sqrt{\frac{2}{d}}\sum_{q=0}^{N} \sin ({k_q x}) \hat{\psi}_{q} (k_q),
\end{align}
where we define
$\psi_{n} (x)=(u_{n\uparrow}(x),u_{n\downarrow}(x),v_{n\uparrow}(x),v_{n\downarrow}(x))$,
and $\hat{\psi}_{q} =(\hat{u}_{q\uparrow},\hat{u}_{q\downarrow},\hat{v}_{q\uparrow},\hat{v}_{q\downarrow})$.
The wavevector $k_q = q \pi/d$ is discretized in terms of the
system width $d$, 
taken to be large enough so
that  the results become independent of $d$.
The next step involves Fourier transforming the real-space BdG equations (Eq.~(\ref{bogo})),
resulting in the following set of coupled equations in momentum space:
\begin{align}
\label{kbogo}
\begin{pmatrix} 
\hat{H}_0 -\hat{h}_z&-\hat{h}_x+i\hat{h}_y&0&\hat{\Delta} \\
-\hat{h}_x-i\hat{h}_y&\hat{H}_0 +h_z&\hat{\Delta}&0 \\
0&{\hat\Delta}^*&-(\hat{H}_0 -\hat{h}_z)&-\hat{h}_x-i\hat{h}_y \\
{\hat \Delta}^*&0&-\hat{h}_x+i\hat{h}_y&-(\hat{H}_0+\hat{h}_z) \\
\end{pmatrix}
\begin{pmatrix}
\hat{u}_{\uparrow}\\\hat{u}_{\downarrow}\\\hat{v}_{\uparrow}\\\hat{v}_{\downarrow}
\end{pmatrix}
=\epsilon_n
\begin{pmatrix}
\hat{u}_{\uparrow}\\\hat{u}_{\downarrow}\\\hat{v}_{\uparrow}\\\hat{v}_{\downarrow}
\end{pmatrix}.
\end{align}
Here we have defined $\hat{u}_\sigma=(\hat{u}_{1\sigma},\hat{u}_{2\sigma},\ldots, \hat{u}_{N\sigma})$,
$\hat{v}_\sigma=(\hat{v}_{1\sigma},\hat{v}_{2\sigma},\ldots, \hat{v}_{N\sigma})$,
and the matrix elements,
\begin{align}
\label{hok}
\hat{H}_0(q,q')&=\frac{2}{d} \int_{0}^d dx \left(\frac{k_q^2}{2m}+\epsilon_\perp-\mu\right) \sin(k_q x) \sin(k_{q'} x),  \\
\label{delk}
\hat{\Delta}({q,q'}) &= \frac{2}{d}\int_{0}^d dx \Delta(x) \sin(k_q x) \sin(k_{q'} x),  \\
\label{hk}
\hat{h}_{i}({q,q'}) &=\frac{2}{d} \int_{0}^d dx\, h_{i}(x) \sin(k_q x) \sin(k_{q'} x), \quad i=x,y,z.
\end{align}
Our numerical procedure for calculating the supercurrent involves
assuming a constant amplitude and phase
for the pair potential
in each $S$ layer, thus
providing  the physically necessary source or sink
of current, via the external electrodes.
We then expand 
the pair potential via Eq.~(\ref{delk}).
Similarly the exchange field and free particle Hamiltonian are
expanded using Eq.~(\ref{hk}) and Eq.~(\ref{hok}) respectively.
We then find the quasiparticle energies and amplitudes 
by
diagonalizing the resultant momentum-space matrix (Eq.~(\ref{kbogo})).
Once the momentum-space wavefunctions and energies are found,
they are transformed back into real-space via Eq.~(\ref{basis}), and
the currents and magnetic moments are calculated as described in Sec.~\ref{sec:theory}.

As mentioned above, the current source arises from the
non self-consistent region where we take $\Delta(x)$ to
be a piecewise constant with  prescribed macroscopic phases in the $S$ electrodes.
The widths of the two outer $S$ terminals
are sufficiently large ($d_S\gg\xi_0$) so that the system boundaries
have a negligible influence on the results.
By taking the
divergence of the current in Eq.~(\ref{cur}) and using the BdG
equations (Eq.~(\ref{bogo})), we find,
\begin{align}
\label{source}
\frac{\partial J_x (x)}{\partial x} =
 2e {\rm Im}\left\{\Delta({x}) \hspace{-.07cm} \sum_n\left[u_{n \uparrow}^*
v_{n\downarrow}+u_{n\downarrow}^*v_{n\uparrow}\right] \
\tanh\left(\frac{\epsilon_n}{2T}\right) \right\},
\end{align}
where the terms in the summation constitute the usual self-consistency equation~\cite{Gennes}
for $\Delta(x)$.
Thus, only when self-consistency 
 in $\Delta({x})$ is achieved, does the right hand side of Eq.~(\ref{source}) vanish, and
current is conserved. 
If the self-consistency condition is not strictly satisfied,
the terms on the right  act effectively as sources of current,
except of course within 
 the ferromagnet regions,  where $\Delta({x})=0$.

\section{Spin current and spin transfer torque}

The spin current can be found
by using the Heisenberg picture.
First we
determine the time evolution of the spin density, ${\bm \eta}({x})$,
\begin{align} \label{scom}
\frac{\partial}{\partial t} \langle {\bm \eta}({x}) \rangle = i \Big\langle [{\cal H},{\bm \eta}({x})] \Big\rangle,
\end{align}
where
the effective BCS Hamiltonian ${\cal H}$ is written,
\begin{align} \label{heff}
{\cal H} =&\int dx \Bigl\lbrace \psi^\dagger ({x})[  {\cal
H}_0({x})-  {\bm h}({x}) \cdot {\bm \sigma}] \psi ({x})  \nonumber \\
 &+
 \Delta({x}) \psi^\dagger_\uparrow({x}) \psi^\dagger_\downarrow({x})
+\Delta^*({x}) \psi_\downarrow({x}) \psi_\uparrow({x})
\Bigr\rbrace.
\end{align}
Here we define,
$\psi({x}) =
(\psi_\uparrow({x}),\psi_\downarrow({x}))^\mathbb{T}$,
and
the
spin density operator  ${\bm \eta}({x})$:
${\bm \eta}({x})  = \psi^\dagger({x})  {\bsigma} \psi({x})$.
Inserting the Hamiltonian (Eq.~(\ref{heff})), 
into Eq.~(\ref{scom}), we end up with
the spin continuity equation:
\begin{align} \label{scon}
\frac{\partial}{\partial t} \langle {\bm \eta}({x}) \rangle +
\frac{\partial}{\partial x} {\bm S}({x}) &=
{\bm \tau}({x}), 
\end{align}
where the spin 
transfer torque ${\bm \tau}$ 
is written in terms of the  expectation value,  
${\bm \tau}(x)= 
2 \langle 
\psi^\dagger({x}) [{\bm \sigma}{\bm \times}{\bm h}] \psi ({x}) \rangle$.
Using the fact that
the spin density is simply related to the magnetization ${\bm m}$ via
${\bm m}({x})  =-\mu_B\, \langle {\bm \eta}({x}) \rangle$,
we end up with Eq.~(\ref{tau1}).
Similarly, the spin current ${\bm S}$ is given by:
\begin{align}
{\bm S}({x}) &= -\frac{i}{2m} \Bigg \langle
\psi^\dagger({x}) {\bm \sigma} \left({\frac{\partial}{\partial x}}\psi({x})\right) - \left({\frac{\partial}{\partial x}}\psi^\dagger({x})\right) {\bm \sigma}
\psi({x}) \Bigg \rangle.
\end{align}
Lastly, we insert the Bogolibuov transformations,
$\psi_{\uparrow}({x})=\sum_n (u_{n\uparrow}({x})\gamma_n - v^*_{n\uparrow}({x})\gamma_n^\dagger)$,  
and 
$\psi_{\downarrow}({x})=\sum_n (u_{n\downarrow}({x})\gamma_n + v^*_{n\downarrow}({x})\gamma_n^\dagger)$,
and use conventional rules for the thermal averages: $\langle \gamma_n^\dagger \gamma_m \rangle = \delta_{nm} f_n$,
$\langle \gamma_m \gamma^\dagger_n \rangle = \delta_{nm} (1-f_n)$, and
$\langle \gamma_n \gamma_m \rangle =  0$,
to 
arrive at Eqs.~(\ref{sx})-(\ref{sz}).
Note that 
${S}_{\sigma}(x)$
represents the spin current flow
along the $x$ direction in
configuration-space, 
with indices $\sigma =
x,y,z$ in  spin-space.

\end{document}